  \def   \ni {\noindent}

  \def   \ssk {\vskip  5truept}
  
  \def   \bsk {\vskip 15truept}

  \def   \newline {\hfil\break}

  \input psfig.tex
  \magnification=1000
  \hsize 5truein
  \vsize 8truein
  \font\abstract=cmr8
  
  \font\text=cmr10     
  \font\affiliation=cmssi10
  \font\author=cmss10
  
  \font\title=cmssbx10 scaled\magstep2

  \def\ref{\par\noindent\hangindent 15pt}
  \nopagenumbers
  \null
  \vskip 3.0truecm
  \baselineskip = 12pt
  
  {\title                       
  \ni Igloo Pixelizations of the Sky
  }                                                             
  \bsk \bsk
  {\author                              
  \ni  Robert G. Crittenden
  }                                                             
  \bsk
  {\affiliation                   
  \ni DAMTP, University of Cambridge, Silver Street, Cambridge CB3 9EW, UK 
\hskip 1.5cm \break
 \ni and CITA, 60 St. George Street, Toronto, ONT M5S 3H8
  }                                                             
  \bsk
  \bsk
  \baselineskip = 9pt
  {\abstract                                                
  \ni 
Upcoming microwave background experiments will see an incredible 
increase in the volume of data to be analyzed, which makes the 
choice of how it is discretized on the sky a crucial issue.
I discuss criteria for evaluating different pixelizations and 
the advantages of using an exactly azimuthal 
or `igloo' pixelization of the sky.  Talk given at 
``The CMB and the Planck Mission" workshop in Santander, Spain, 
June 1998. 

  }                                                              
  \bsk
  \baselineskip = 12pt
  {\text                                                       
  \ni 1. WHY DO WE CARE ABOUT PIXELIZATION?

The next generation of cosmic microwave experiments will usher in a new 
`precision' era, providing maps of large parts of the microwave sky 
with greater resolution than ever before.   
In particular, the satellite experiments, MAP and Planck, will provide millions or 
hundreds of millions of independent temperature measurements over the full sky.  
With such a dramatic increase in the amount of data, the methods that we use 
to handle and analyze it must be substantially improved.  
The first step in this process is determining how the data are to 
be discretized and stored.  

The only previous full sky measurement of the CMB anisotropies was COBE DMR, 
which had very low resolution and so had only a few thousand independent 
pixels.  The COBE data was stored using a `quad-cubed' pixelization, 
which was based on the edges of a cube that were projected onto the sky. 
Each face of the cube was divided into quarters of the same area, and this 
process was iterated to obtain higher and higher resolution pixelizations. 
(In the COBE implementation, however, the pixels were not strictly equal area 
and had variations in pixel size of order 10\% [Greisen \& Calabretta 1993].)   
Most analyses of the 
COBE data used brute force inversions of $N \times N$ matrices, where 
$N$ is the number of pixels.
The time required to perform such inversions scales as $N^3$, so while this 
was just tractable for COBE, it will not be an alternative for the future satellite 
experiments.  

Many alternative pixelizations have been suggested, each with its own 
advantages [Tegmark 1997, Gorski 1997, Wright 1998, Crittenden \& Turok 1998].  
When choosing between them, a number of issues must be 
considered: 

\ni $\bullet$ Speed of Spherical Transforms -- One often wants to 
change variables from the experimentally measured pixel temperatures ($T_P$)
to the coefficients of a spherical harmonic expansion 
($a_{lm}$'s) to compare 
to predictions of different theories.   
These are related by a spherical transform, defined by 
$$ T(\theta,\phi) = \sum_{l,m} a_{lm} Y_{lm}(\theta,\phi), $$
$$ a_{lm} = \int T(\theta,\phi) Y^*_{lm}(\theta,\phi) d\Omega. $$ 
When discretized, these transforms naively take $N^2$ operations, 
because $N$ spherical harmonic functions need to be evaluated 
at $N$ separate points on the sky.  

However, as has been recent emphasized, if one uses a pixelization
with discrete azimuthal symmetry, then the spherical transforms can be greatly 
sped up [Muciaccia, Natoli \& Vittorio 1997].  
This utilizes the fact that the azimuthal dependence of the 
spherical harmonic functions can be simply factored out, 
$$ Y_{lm}(\theta,\phi) = \lambda_{lm}(\theta) e^{im\phi}. $$  
The azimuthal sum can then be performed quickly with a fast Fourier transform. 
Effectively, this means that the $N$ functions need only be evaluated 
at $N^{1/2}$ different latitudes, so that the whole process requires 
only $N^{3/2}$ operations.  
This property has recently been exploited by Oh, Spergel and Hinshaw [1998] to 
solve for the power spectrum of simulated MAP data.   

\ni $\bullet$ Convenience --  One of the nice features of the 
quad-cubed pixelization used by COBE was its hierarchical nature, 
in that each pixelization was a subdivision of a coarser pixelization. 
This naturally gave the pixels a tree structure and allowed the data 
to be coarsened by simply adding the finer resolution temperatures in 
groups of four.   Another benefit of this tree structure was that it 
ordered the data in a local way, which allowed for quick algorithms 
for finding neighboring pixels.  It would very useful to maintain this 
hierarchical structure in future pixelizations. 

\ni $\bullet$ Simplicity -- The pixelization should be easy to understand,
use and explain.
An underlying simplicity helps also to make the algorithms for manipulating 
the data faster. 

\ni $\bullet$ Systematic Effects --  Finally, because one is making a 
precision measurement, one needs to minimize systematic errors 
introduced by the pixelization and be able to correct for them.  
The pixelizing of the sky creates two kinds of errors 
as one approaches the pixel scale.  
First, the pixelization smoothes out the temperature maps, effectively 
suppressing the amplitudes of the different modes.  The suppression 
becomes larger for the higher $l$ modes, and it also depends on the 
individual $m$ modes.   This damping must be understood and accounted 
for in order to reconstruct the original mode amplitudes. 

Secondly, one must consider the problem of aliasing. 
Aliasing arises because the different modes cease to be orthogonal when they 
are discretized. 
Clearly, one can not reconstruct any more independent mode amplitudes than the 
number of independent pixels. 
For a regular one dimensional grid, only modes with a frequency below the 
characteristic frequency of the grid, known as its Nyquist frequency, are 
orthogonal.  Higher frequency modes 
appear the same as these lower modes when discretized, 
which is demonstrated by Figure 1.  

For regular grids, 
modes below the Nyquist frequency remain orthogonal when they are discretized. 
However, this is not true for pixelizations of the sphere, which complicates 
the inversion.  
This also leads to some ambiguity about which modes are chosen 
to be extracted, but the natural choice remains the $N$ longest wavelength
modes, since 
beam profile damps the higher $l$ mode amplitudes.  
The orthogonality of the lowest modes is best preserved by making the
pixels as round as possible.   
To quantitatively correct for the remaining lack of orthogonality, it is 
also important to be able to exactly integrate these modes over 
the pixelization. 

  }
 \midinsert
 \vskip -0.5truecm
 \par\noindent
 \centerline{\psfig{figure=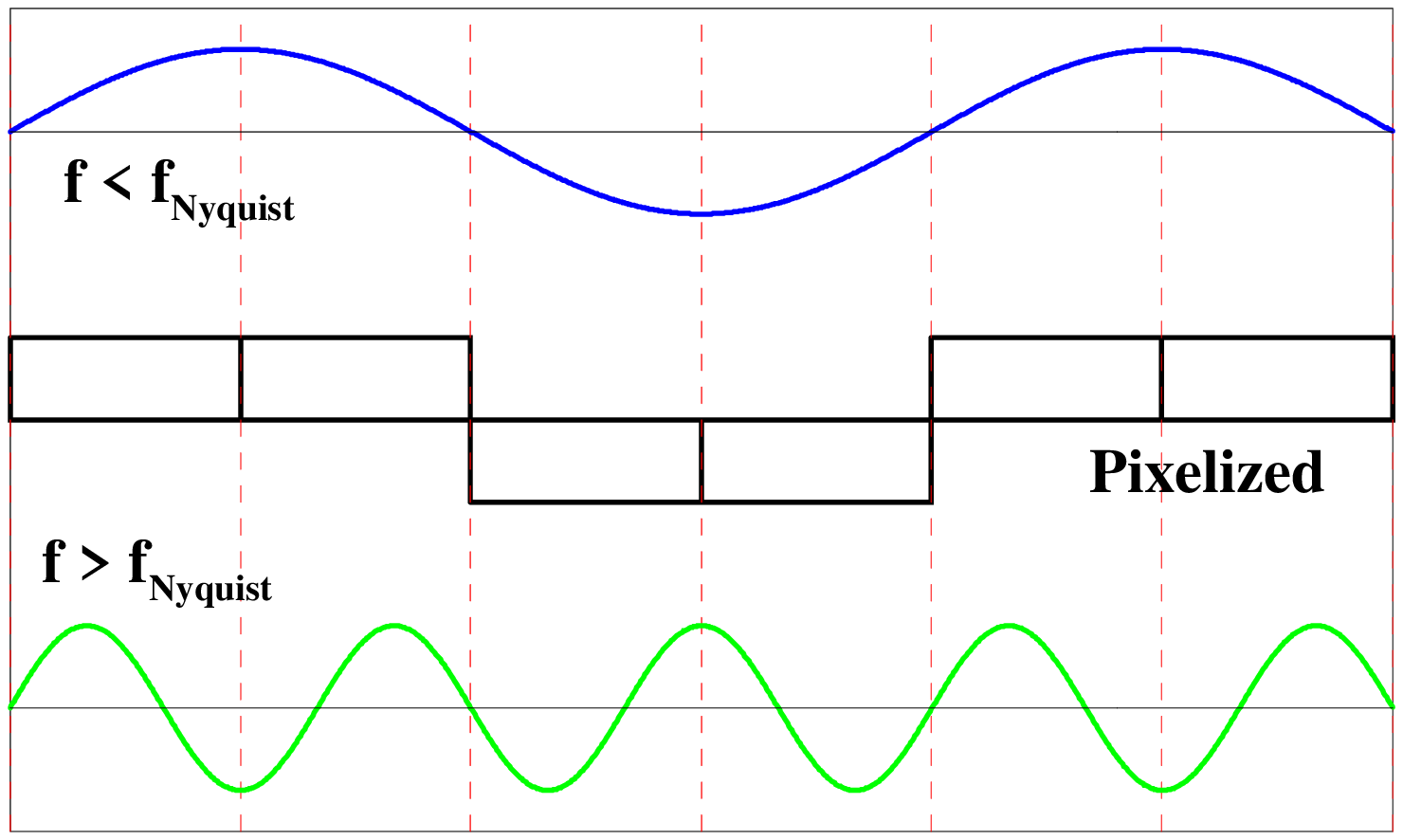,width=12truecm}}
 
\leftskip=0.5truecm
	\rightskip=0.5truecm
		  \vskip -3truecm
\noindent
{\abstract FIGURE 1. Aliasing in 1-dimension -- modes with frequencies higher 
than the Nyquist frequency can appear the same as those below it. 
   }
	  \vskip 0.5truecm
  \leftskip=0truecm
  \rightskip=0truecm
  \endinsert

  {\text 
  \ni 2. IGLOO PIXELIZATIONS

Perhaps the simplest pixelizations which satisfy all of these requirements 
is the class we refer to as `igloo' pixelizations.   By this, we mean any 
division of the sphere into layers perpendicular to some axis, 
where each layer is divided into 
identical pixels with discrete azimuthal symmetry. 
These have the advantage that they are naturally azimuthal, so that using 
the FFT is exact.   
One example of an igloo pixelization can be seen in the design of the 
spider web bolometers for Planck.  

Another simple example of an igloo pixelization is to divide the sky into 
equal divisions of latitude and longitude.  (This is sometimes known as an 
Equidistant Cylindrical Projection, or ECP.)  
The difficulty of this is that it leads to a range of pixel sizes, with pixels 
near the poles becoming smaller and thinner than those at the equator.  
This means some pixels are effectively wasted.  To avoid this, we can group the 
pixels together as we approach the poles, which allows us to keep the pixels 
approximately the same area.  

Igloo pixelizations are not hierarchical in general, but can be made so easily.
To do this, we begin with an igloo pixelization which has been designed 
to minimize the pixel distortion.  We then can subdivide each pixel in 
$\theta$ and $\phi$, to create a pixelization with four times as many pixels. 
The pixels which touch a pole are divided into one polar pixel, surrounded 
by three other pixels.  (See Figure 2.)

With more pixels in the base pixelization, the pixels can be made less distorted, so 
there is a tradeoff between minimizing pixel distortion 
and making the scheme as simple and hierarchical as possible. 
 One simple example is to begin with twelve base pixels, three at each 
pole and six around the equator.
(For simplicity, we refer to this as a 3:6:3 pixelization.) 
Each base pixel is roughly 
$60^{\circ}\times 60^{\circ}$.  Another possibility we have considered is to 
divide the sky into 12,000 base pixels, approximately $1^{\circ}\times 1^{\circ}$.
These pixels are more uniform than those in a 3:6:3 pixlization of the same 
resolution.  (Further discussion of these pixelizations can be found in 
Crittenden \& Turok [1998].)

Thus, igloo pixelizations are naturally azimuthal, 
and can simply be made hierarchical with fairly low pixel distortion. 
They also have the advantage that the pixel edges are simple 
to specify, as they are lines of constant latitude or longitude. 
This makes it possible to integrate spherical harmonic functions 
over the pixels quickly.  (The integral factorizes, and can be 
evaluated recursively.)  This is essential to simulate the experiments
and to understand the effects of aliasing of the pixelization.  
  }
 \midinsert
 \vskip -0.5truecm
 \par\noindent
 \centerline{ 
\psfig{figure=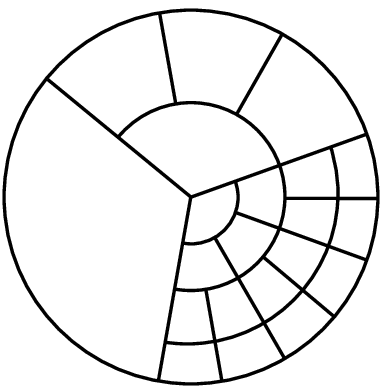,width=5truecm} 
  \psfig{figure=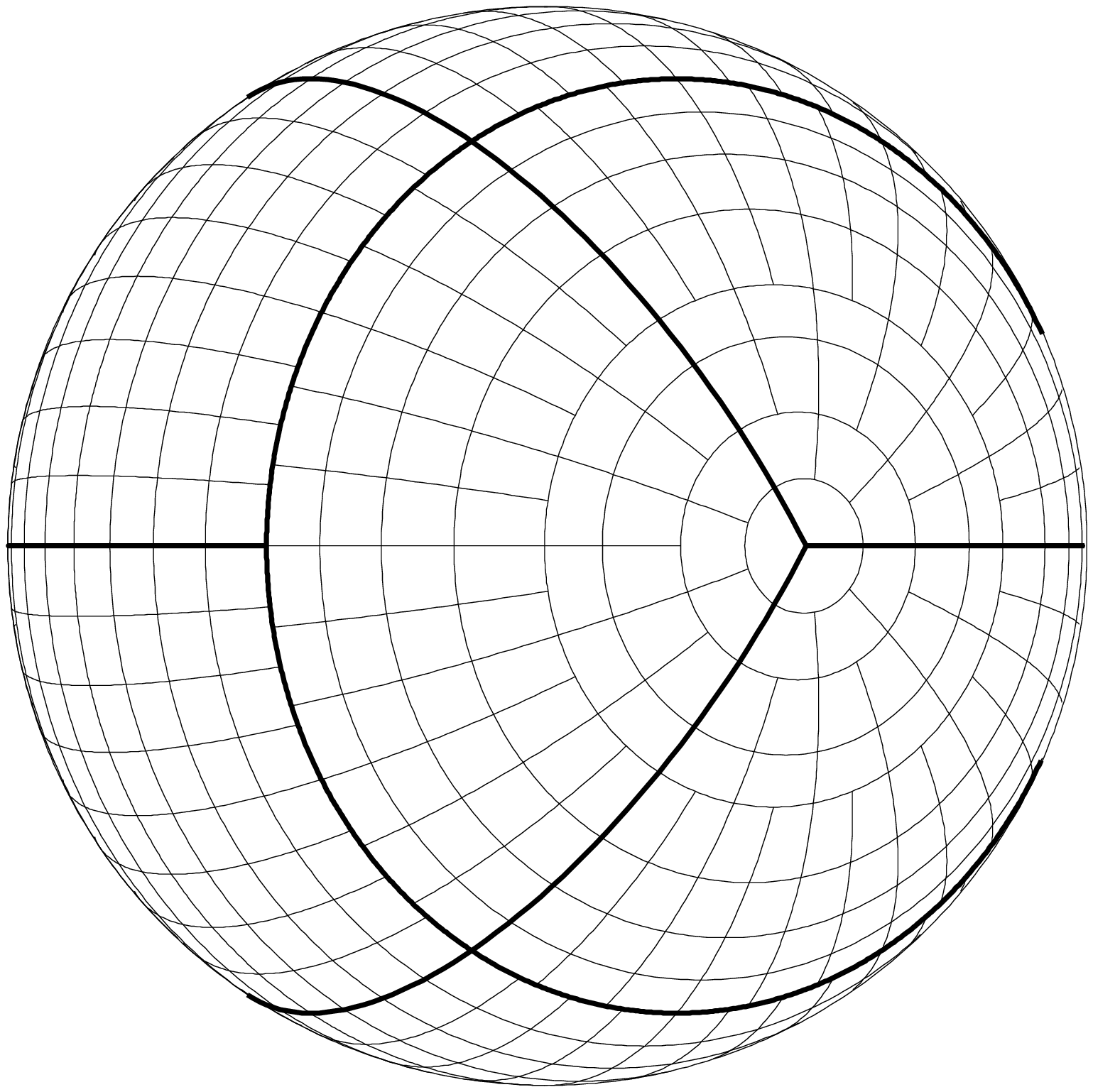,width=7truecm}}
 
\leftskip=0.5truecm
	\rightskip=0.5truecm
		  \vskip -1truecm
\noindent
{\abstract FIGURE 2. The left figure shows a polar cap division scheme which is 
hierarchical and causes little pixel distortion. 
This is implemented in the right figure, a 3:6:3 pixelization, 
with each of its twelve base 
pixels broken into 64 subpixels in an equal area way.  
   }
	  \vskip 0.5truecm
  \leftskip=0truecm
  \rightskip=0truecm
  \endinsert

  {\text 
  \ni 3. EVALUATING PIXELIZATIONS

Given two pixelizations, how does one judge between them?  This is difficult to 
determine naively, because the pixelization is used for such a wide variety of 
purposes, from the initial mapping of the beam all the way through to the final 
analysis to obtain parameter constraints.  The ideal test would be 
to simulate the whole process from beginning to end, but this is difficult 
to do in practice.  Here we focus on one of the most important issues -- 
given a continuous 
temperature map, what is the effect of the pixelization on the ability to 
reconstruct its initial power spectrum?  

Other suggested pixelization criteria have 
focused on shape distortions of the individual pixels or on aliasing of 
the monopole alone into other modes [Wright 1998, Tegmark 1997].
While these are important issues, they only partially address the 
question. 
A better test is to create realistic 
simulated maps and see how well one can invert them to find 
the initial power spectrum.  This factors in all 
of the effects of aliasing and mode suppression discussed above.   

Consider first the effect of discretizing a single mode, $Y_{lm}(\theta,\phi)$. 
The temperature of this mode, averaged over each pixel $P$, can be written 
as $W_{lm}^P = \int_PY_{lm}(\theta,\phi) d\Omega$.  Since the continuous modes 
are normalized to unity, the suppression due to pixelization -- effectively 
its window function -- is given by the norm of the 
pixelized mode, $N_{lm} = \sum_P A_P W_{lm}^P W_{lm}^{*P} $. 
Ideally, this function would be one for the lowest $N$ modes, and then drop off 
to zero for the higher modes.  Realistically, this response drops off 
more slowly, at a scale determined by the the effective average area of the pixels. 

The pixelized temperature map can be simply expanded in terms of these discrete 
functions, 
$T_P = \sum_{lm} a_{lm} W_{lm}^P$.  Like the continuous map, the pixelized 
map can be expanded in terms of spherical harmonics, with coefficients 
$a_{lm}^{pix}$.  
These coefficients are given by the transform of the pixel temperatures and are 
a linear combination of the original, continuous $a_{lm}$'s, 
$$ a_{lm}^{pix} = \sum_P A_P W_{lm}^{P*} T_P \equiv {\cal M}_{lml'm'}a_{l'm'}.$$
For modes which vary slowly compared to the pixel size, this coefficient matrix is 
very nearly diagonal, ${\cal M}_{lml'm'} \simeq N_{lm}\delta_{ll'}\delta_{mm'}$. 
This is not the case for modes close to the pixel size, where there is 
strong cross talk between the modes.  This matrix is not invertible, because 
there are many more degrees of freedom in the continuous map than the 
pixelized one.  We can try to invert if we assume some constraint on the 
$a_{lm}$'s, for example that they are zero above some cutoff, and this allows 
a reasonable recovery of most of the modes, assuming that modes below the 
pixel scale are suppressed (such as by the beam smearing.) 
This inversion however is time consuming and often is very slow to converge, 
particularly for modes that are nearly degenerate when pixelized. 

For many applications, we are only interested in the power spectrum 
rather than in amplitudes of particular modes.   Similarly to the above 
analysis, 
we can relate the true spectrum to the measured pixelized spectrum, defined by
$C_l^{pix} = \sum_m |a^{pix}_{lm}|^2/(2l+1)$. 
Naively dividing by the suppression of each mode to estimate the true 
power spectrum overestimates the power, because of the lack of orthogonality 
of the modes.
However, assuming rotational invariance, the expectation 
value of the pixelized power spectrum is related to the true one by 
$\langle C_l^{pix}\rangle = {\cal M}_{ll'}C_{l'}$, where ${\cal M}_{ll'}$ is 
a matrix which depends solely on the pixelization.  Thus, one can make an 
unbiased estimate of the true power spectrum: 
$$C_{l'}^{unbiased} = {\cal M}_{ll'}^{-1} C_l^{pix}.$$
The effect of using this corrected estimator is shown in figure 3.
The power spectrum recovery at large $l$ is substantially improved
 [Crittenden \& Turok 1998]. 

Finally, if the sky is pixelized at much higher resolution than the 
beam size then the systematic errors we have considered here will 
be small.  However, doing this would require many times more pixels,  
with a hefty cost in computation time.  In practice, we will want to use as 
low resolution pixelization as possible, so that understanding these 
systematic errors is essential.  

  }
 \midinsert
 \vskip 0.1truecm
 \par\noindent
 \centerline{\psfig{figure=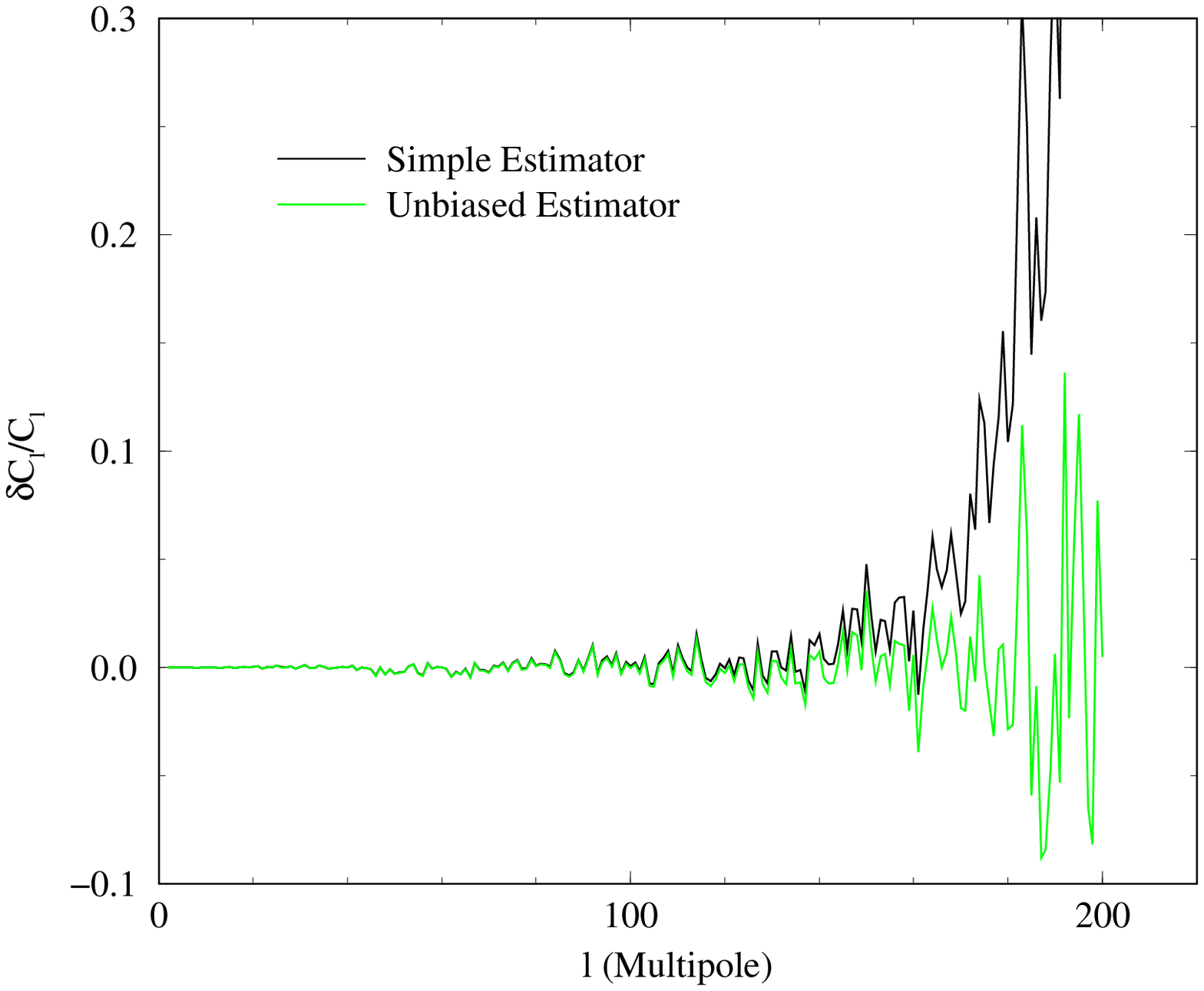,width=8truecm}}
 
\leftskip=0.5truecm
	\rightskip=0.5truecm
		  \vskip 0.5truecm
\noindent
{\abstract FIGURE 3. This shows the improvement in the power spectrum estimation 
using the unbiased method described above.  The simple estimator, where each mode is 
simply divided by its pixelized normalization, consistently overestimates the power. 
(Based on 50,000 pixels, $l_{Nyquist} = 192$.) 
   }
	  \vskip 0.5truecm
  \leftskip=0truecm
  \rightskip=0truecm
  \endinsert

  {\text 
  \ni 4. CONCLUSIONS

The choice of pixelization is a crucial one.  It affects the speed and ease of 
analysis, and introduces systematic effects for modes of the scale of the pixels. 
These errors are largest precisely where cosmic variance is smallest, for the very 
high $l$'s.
Minimizing these systematic effects requires the pixels to be as uniform and 
circular as possible.  Correcting for them requires a full understanding of 
aliasing and suppression of the modes, and so the ability to integrate 
over the pixels.  

Igloo pixelizations are well suited for this purpose.  They are simple, 
maximally azimuthal 
and can easily be made hierarchical with   
quite uniform pixels.  In addition, they are quickly integrable, 
which allows us to correct for aliasing and suppression effects. 

Software for CMB map making and inversion in the igloo pixelization can be found at: 
{\tt http://www.damtp.cam.ac.uk/user/rgc1002/pixel.html}
I would like to thank the organizers of the Santander conference and
would especially like to acknowledge my collaborator on this work, Neil Turok. 
 }
   
 %
  
  
  \vskip 0.1truecm
  {\text 
  \ni {REFERENCES}
  \ssk
  
  \ref Crittenden \ R.G.. \& Turok N.G., 1998, submitted {\sl Ap.J.}, 
astro-ph/9806374
  \ref Gorski K., 1997, private communication
  \ref Greisen E.W. \& Calabretta M., 1993, {\sl Bull. American Astron. Soc.} {\bf 182} 09.01
  \ref Muciaccia \ P.F., Natoli P. \& Vittorio N., 1997, {\sl Ap.J.}, {\bf 488}, L63
  \ref Oh \ S.P., Spergel D.N. \& Hinshaw G., astro-ph/9805339
  \ref Wright, E., 1998, private communication
  }                                                            
  
  \end